\documentclass[sigchi]{acmart}

\copyrightyear{2020}
\acmYear{2020}
\setcopyright{acmlicensed}
\acmConference[TEI '20]{Fourteenth International Conference on Tangible, Embedded, and Embodied Interaction}{February 9--12, 2020}{Sydney, NSW, Australia}
\acmBooktitle{Fourteenth International Conference on Tangible, Embedded, and Embodied Interaction (TEI '20), February 9--12, 2020, Sydney, NSW, Australia}
\acmPrice{15.00}
\acmDOI{10.1145/3374920.3374941}
\acmISBN{978-1-4503-6107-1/20/02}

\usepackage{multicol}
\usepackage{cuted}
\usepackage[font=small,labelfont=bf,tableposition=top]{caption}
\usepackage{enumitem}
\usepackage{balance}
\newcommand {\changes}[1]{{#1}}

\newcommand{\tool}{LiftTiles}

\begin{document}
% \title{\tool{}: Modular and Reconfigurable Room-scale Shape Displays through Retractable Inflatable Actuators}
\title{\tool{}: Constructive Building Blocks for Prototyping Room-scale Shape-changing Interfaces}

\author{Ryo Suzuki}
\affiliation{University of Colorado Boulder}
\email{ryo.suzuki@colorado.edu}

\author{Ryosuke Nakayama}
\affiliation{Keio University}
\email{ryosk611@sfc.keio.ac.jp}

\author{Dan Liu}
\affiliation{University of Colorado Boulder}
\email{dali2731@colorado.edu}

\author{Yasuaki Kakehi}
\affiliation{The University of Tokyo}
\email{kakehi@iii.u-tokyo.ac.jp}

\author{Mark D. Gross}
\affiliation{University of Colorado Boulder}
\email{mdgross@colorado.edu}

\author{Daniel Leithinger}
\affiliation{University of Colorado Boulder}
\email{daniel.leithinger@colorado.edu}

\begin{abstract}
Large-scale shape-changing interfaces have great potential, but creating such systems requires substantial time, cost, space, and efforts, which hinders the research community to explore interactions beyond the scale of human hands. We introduce modular inflatable actuators as building blocks for prototyping {\it room-scale} shape-changing interfaces.  Each actuator can change its height from 15cm to 150cm, actuated and controlled by air pressure. Each unit is low-cost (8 USD), lightweight (10 kg), compact (15 cm), and robust, making it well-suited for prototyping room-scale shape transformations. Moreover, our modular and reconfigurable design allows researchers and designers to quickly construct different geometries and to explore various applications. This paper contributes to the design and implementation of highly extendable inflatable actuators, and demonstrates a range of scenarios that can leverage this modular building block.
\end{abstract}

\begin{CCSXML}
<ccs2012>
<concept>
<concept_id>10003120.10003121</concept_id>
<concept_desc>Human-centered computing~Human computer interaction (HCI)</concept_desc>
<concept_significance>500</concept_significance>
</concept>
</ccs2012>
\end{CCSXML}

\ccsdesc[500]{Human-centered computing~Human computer interaction (HCI)}
\keywords{inflatables, shape-changing interfaces, large-scale interactions}

% This paper proposes constructive building blocks for prototyping room-scale shape-changing interfaces.
% due to the technical challenges in scalability, robustness, and deployability.
% of reconfigurable room-scale shape-changing interfaces and discuss how this prototype environment can open up the new opportunity, 

% and light (e.g., 1.8kg), while extending up to 1.5m to allow for large-scale shape transformation. Inflatable actuation also provides a robust structure that can support heavy objects. Due to its modular design, users can easily deploy and rearrange Dynaroom units for different applications. This paper describes the design and implementation of Dynaroom and explores the application space 
% Dynaroom consists of an array of retractable and inflatable actuator that is compact (e.g., 15cm tall) and light (e.g., 1.8kg), while extending up to 1.5m to allow for large-scale shape transformation. Inflatable actuation also provides a robust structure  This paper describes the design and implementation of Dynaroom and explores the application space for reconfigurable room-scale shape displays. 
% a modular and reconfigurable room-scale shape display. Shape displays have been a promising approach to general-purpose shape-changing interfaces. However, most of the existing pin-based shape displays focus on interactions at the scale of a human hand, and due to , few explore larger-scale shape transformations. To overcome these challenges, this paper proposes 

\begin{teaserfigure}
% \vspace{-0.2cm}
\centering
\includegraphics[width=1\textwidth]{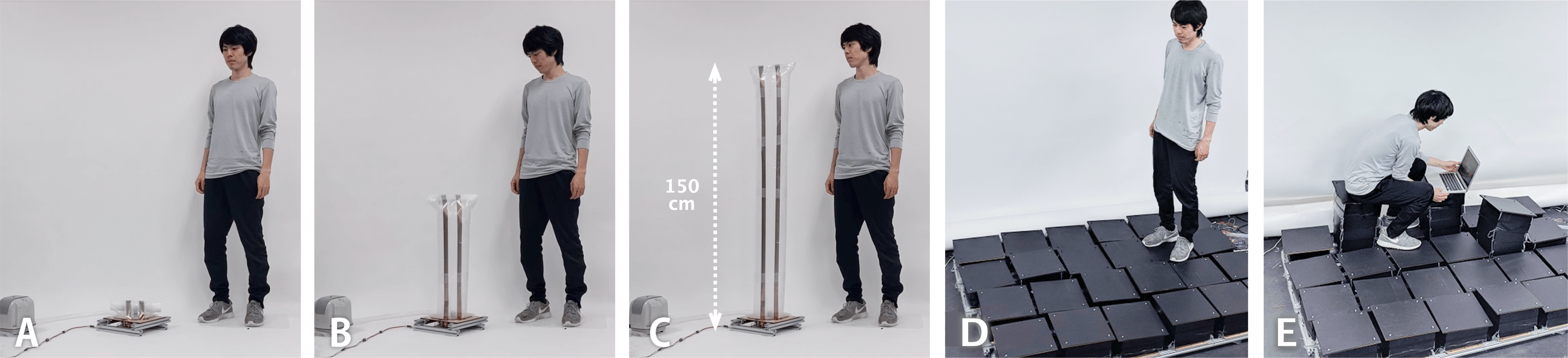}
\caption{A modular inflatable actuator as a building block for prototyping room-scale shape-changing interfaces. Each actuator can extend from 15 cm to 150 cm (A-C). By constructing these modular blocks, a designer can easily prototype different room-scale shape-changing interfaces, such as a 5 x 5 array of room-scale shape display that a user can sit down and step on (D-E).}
~\label{fig:cover}
% \vspace{-0.2cm}
\end{teaserfigure}

\maketitle 

\section{Introduction}
Emerging shape-changing user interfaces have great potential to transform interactions between humans and computers. 
In contrast to traditional graphical user interfaces, shape-changing interfaces provide rich physical affordances and haptic sensations by adapting their shape to the users' needs and context.
While prior work mostly focuses on shape-change of hand-held objects (e.g. shape-changing mobile phones~\cite{gomes2013morephone, nakagaki2015lineform, roudaut2013morphees, yao2013pneui}) or tabletop surfaces (e.g., shape displays~\cite{follmer2013inForm, iwata2001project, poupyrev2004lumen}), larger-scale shape transformation that engages the whole body has been proposed~\cite{green2016architectural, sturdee2015public}.
However, to construct and deploy such large-scale shape-changing interfaces requires substantial time and effort, at often prohibitive cost. These barriers restrict the research community to prototype and explore applications beyond the scale of the human hand. Our work is addressing this challenge by developing more accessible prototyping tools to help researchers, designers, and practitioners to explore the broader space of possible applications.

This paper introduces \tool{}, modular inflatable actuators for prototyping room-scale shape-changing interfaces.
Each inflatable actuator has a large footprint (e.g., 30 cm x 30 cm) and enables large-scale shape transformation.
The actuator is fabricated from a flexible plastic tube and constant force springs. It extends when inflated and retracts by the force of its spring when deflated. By controlling the internal air volume, the actuator can change its height from 15 cm to 150 cm (Figure~\ref{fig:cover} A-C).
We designed each module as low cost (e.g., 8 USD), lightweight (e.g., 1.8kg), and robust (e.g., withstand more than 10 kg weight), so that it is suitable for rapid prototyping of room-sized interfaces.
Inspired by reel-based pneumatic actuators~\cite{hammond2017pneumatic}, our design utilizes constant force springs to provide greater scalability, simplified fabrication, and stronger retraction force, all essential for large-scale shape-change. 
\changes{Building on our prior initial exploration of an inflatable actuator~\cite{suzuki2019lifttiles}, we describe the design rationale, technical details and limitations of our system in detail.}

\tool{} leverages modular and reconfigurable hardware building blocks. Designers can rearrange these blocks into various geometries to quickly explore different form factors and applications.
We demonstrate some prototyping example applications, such as a shape-changing floor for adaptive furniture, a shape-changing wall for dynamic partition and information display, and environmental haptic interfaces for immersive VR experiences.

This paper contributes:
\begin{enumerate}
% \item \tool{}, a proof-of-concept prototype for modular and reconfigurable room-scale shape displays.
\item Concept of a modular inflatable actuator as a building block for prototyping room-scale shape-changing interfaces.
\item Design and implementation of a highly extendable linear actuator that is low-cost, lightweight, compact, robust, and easy-to-fabricate.
\item A range of applications that demonstrate the potential of the system for prototyping room-scale shape transformation. 
\end{enumerate}

\section{Related Work}

\subsection{Inflatable Shape-Changing Interfaces}
Inflatables have long been a popular method for constructing large and lightweight architectural structures and artistic sculptures~\cite{farm1970inflatocookbook, piene2008sky}. 
These objects are usually passive after inflation, but HCI researchers have explored their use for dynamic shape-changing interfaces (e.g PneUI~\cite{yao2013pneui}, aeroMorph~\cite{ou2016aeroMorph}, JamSheets~\cite{ou2014jamsheets}, Printflatables~\cite{sareen2017printflatables}) and active haptic interfaces (e.g., ForceJacket~\cite{delazio2018force}, PuPoP~\cite{teng2018pupop}). 
\changes{Swaminathan et al. also explored room-scale deployable structures with inter-connected pneumatic trusses~\cite{swaminathan2019input}}
One key advantage of inflatables is their ability to transform their size and shape drastically. 
For example, reel-based pneumatic actuators~\cite{hammond2017pneumatic, hawkes2017soft} are highly extendable actuators for soft robots and shape-changing interfaces.
Our key technical contribution is to extend this line of work to a larger footprint (e.g., 30 cm x 30 cm) and enable a more robust structure (e.g., withstand 10 kg weight).
\changes{Most closely related to our work, TilePoP~\cite{teng2019tilepop} introduces a room-scale inflatable shape display for haptics in VR. Compared to the TilePop modules, our actuator employs a constant force spring for continuous, rather than discrete extension. The spring also removes the need for a vacuum pump, as it pushes out air when deflating. We also explore applications besides VR that leverage the general-purpose nature of modular, reconfigurable actuation blocks.}

\subsection{Prototyping Shape-changing Interfaces}
Shape-changing interfaces are technically challenging to build and control. Therefore, creating appropriate prototyping tools for these interfaces is considered one of the most important and critical goals in shape-changing UI research~\cite{alexander2018grand}.
Paper sketches are commonly used to design shape-changing interfaces~\cite{Rasmussen2016SketchingShapeChanging}, but they are static and lack the affordances of physical objects. Tools like ShapeClips~\cite{hardy2015shapeclip} and Kinetic Sketchup~\cite{Parkes2009Sketchup} have demonstrated how rapid physical prototyping tools enable designers to explore a range of applications in hours, rather than weeks.
\changes{For large-scale shape transformation, TrussFormer~\cite{kovacs2018trussformer} introduces a tool to design and prototype a large-scale kinetic transformation by actuating a static truss structure~\cite{kovacs2017trussfab}.
Prototyping tools for large-scale shape transformation could open up exciting application spaces~\cite{sturdee2015public}, but researchers still lack access to such tools. Our work contributes to this by introducing inexpensive, deployable and reconfigurable building blocks to rapidly prototype large-scale, functional shape-changing interfaces. We also propose example applications built with these building blocks.}
% These tools have demonstrated interactions at the scale of human hands.

\subsection{Large-scale Shape-changing Interfaces}
In the past, large-scale shape-change has been explored primarily in architecture and art.
For example, several large-scale ceiling-mounted or wall-based shape displays have been built (e.g., Kinetic sculpture at the BMW Museum~\cite{fermoso2008kinetic}, Breaking the Surface~\cite{rhodes_2014}, Hyposurface~\cite{goulthorpe2001aegis}, MegaFaces~\cite{khan2014megafaces}).
However, building these large installations requires tremendous time, cost, and efforts.
They are usually designed for a single-purpose installation and difficult to customize, once built.
In contrast, our goal is to provide a reconfigurable kit that supports rapid prototyping.
Some works explore a modular approach to construct dynamic furniture (e.g., Lift-bit~\cite{morillo2013radical}, Tangible Pixels~\cite{tang2011tangible}), but these electro-mechanical modules are expensive when a larger size and stronger force are required, placing them outside the reach of most researchers.
We aim to reduce the required cost and time by one or two orders of magnitude by leveraging low-cost and easy-to-fabricate inflatable actuation methods.

\section{LiftTiles: System Design}

\subsection{Overview}

This section describes the design and implementation of our modular inflatable actuator and how we use it to construct room-scale shape-changing interfaces.

\begin{figure}[!h]
\centering
\includegraphics[width=3.2in]{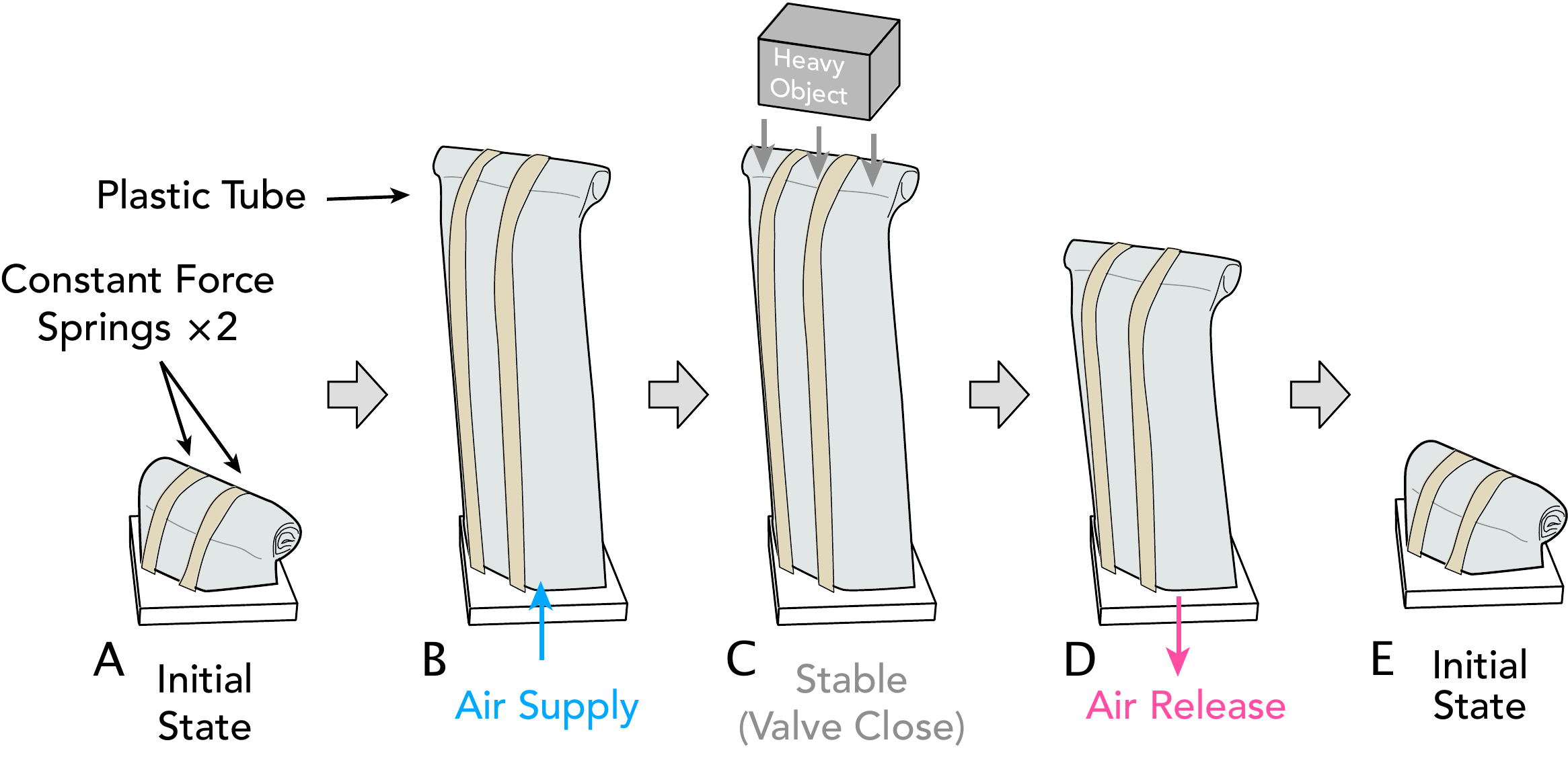}
\caption{Mechanism of our inflatable actuator.}
~\label{fig:system-mechanism}
\vspace{-0.6cm}
\end{figure}

Our inflatable actuator leverages an extendable structure similar to a party horn.
Each inflatable actuator consists of a flexible plastic tube and two constant force springs, which are rolled at their resting positions (Figure~\ref{fig:system-mechanism}).
When pumping air into the tube, the actuator extends as the internal air pressure increases and the end of the tube unwinds. When releasing air through a release valve, the inflatable tube retracts due to the force of the embedded spring returning to its resting position.

\begin{figure}[!h]
\centering
% \includegraphics[width=3.4in]{final-figures/system-cad-illust.jpg}
% \hspace{0.02in}
\includegraphics[width=1.5in]{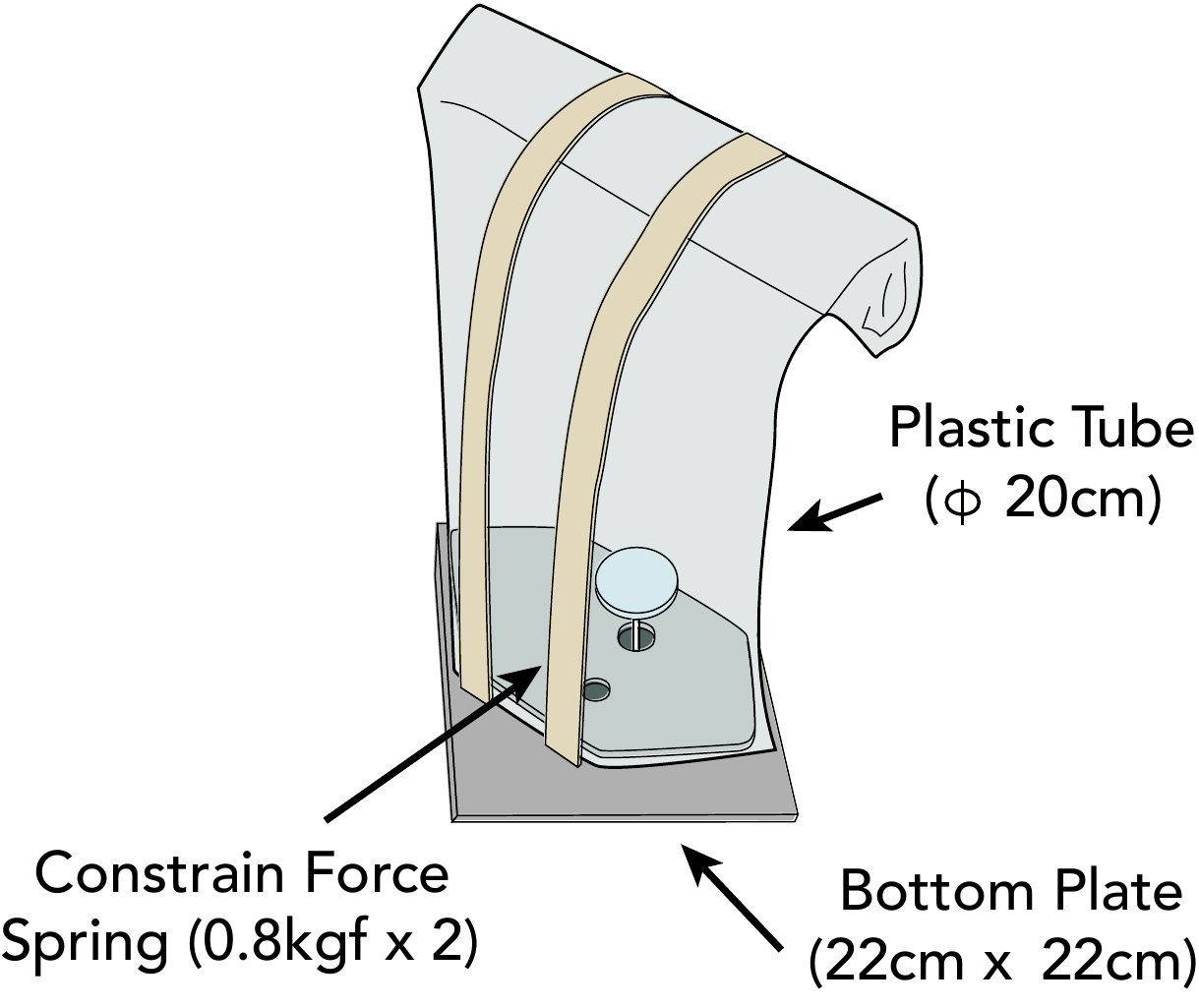}
\hspace{0.04in}
\includegraphics[width=1.65in]{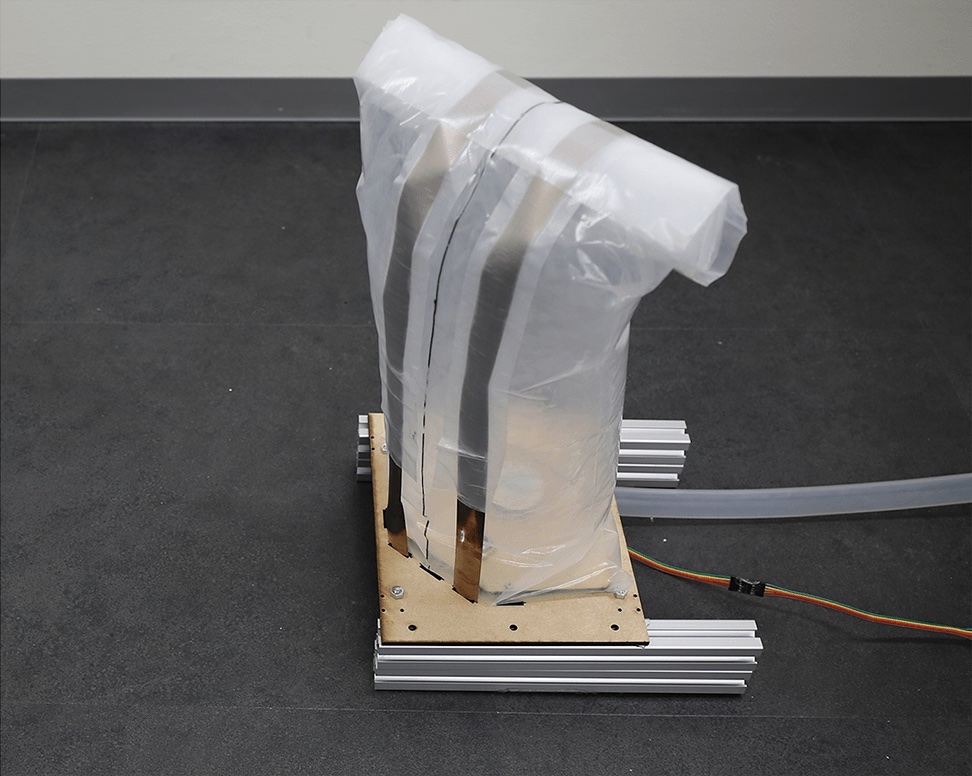}
\caption{Each actuator consists of a plastic tube and constant force springs, which extend from 15 cm to 150cm when inflated and retract when deflated.}
~\label{fig:mechanical-design}
% \vspace{-0.3cm}
\end{figure}

\begin{figure}[!h]
\centering
\includegraphics[width=3.4in]{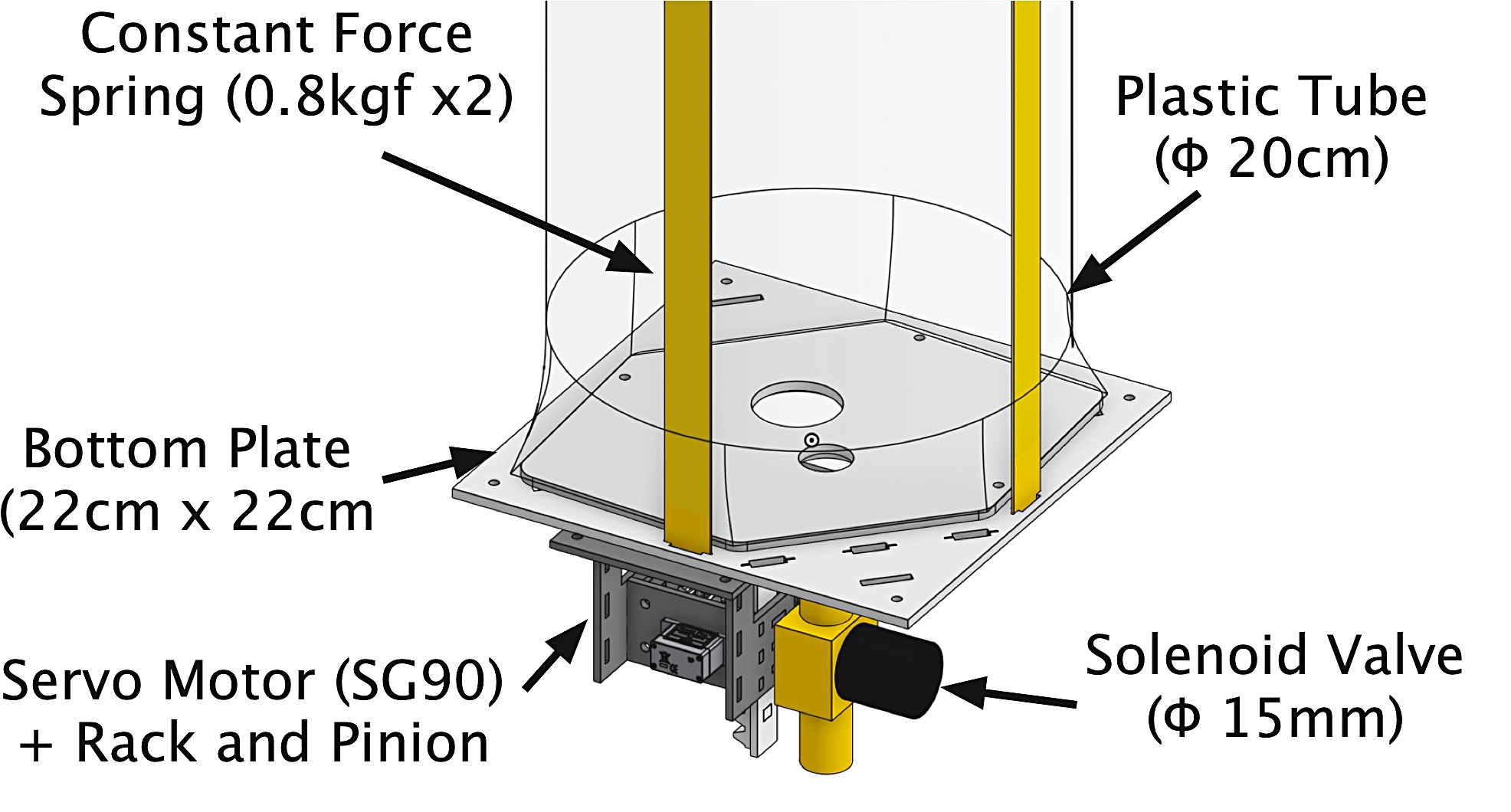}
\caption{3D CAD model of the actuator design, including the inflatable tube, the two attached constant force springs, solenoid valve for air supply, servo motor for air release, and bottom plate to fix the inflatable tube.}
~\label{fig:system-cad}
\vspace{-0.6cm}
\end{figure}

Figure~\ref{fig:mechanical-design} and~\ref{fig:system-cad} illustrates an overview of our actuator design.
Our design is inspired by the prior reel-based actuator~\cite{agrawal2015protopiper, hammond2017pneumatic, hawkes2017soft, suzuki2019shapebots, takei2011kinereels}, which enables the highly extendable structure.
Particularly, Hammond et al.~\cite{hammond2017pneumatic} pioneered the highly extendable reel-based actuator with pneumatic actuation. However, as we built prototypes, we realized that the force of the spiral torsion springs and the 3D printed enclosure used in \cite{hammond2017pneumatic} were too weak to retract a large pneumatic tube and support heavy objects.
Also, the retractable force of the spiral springs varies with the length of the inflatable tube (e.g., weak spring force at shorter extension).
Our design contributes to overcoming these challenges for scalable, reliable, and easy-to-fabricate actuation.

\subsection{Design Rationale}
The goal of our system is to {\it provide an accessible prototyping building block for room-scale shape-changing interfaces}. To achieve this goal, we set the following three design requirements:

\subsubsection{Low-cost and Scalable}
Cost and fabrication complexity is essential to lower the barrier of prototyping.
As we discussed, commercially available mechanical actuators are expensive when the system requires a larger size and higher load resistance. Therefore, we considered inflatables for low-cost and scalable actuation methods.

\subsubsection{Robust}
For room-scale applications, the strength and robustness of the actuator is another important factor.
Unlike table-top interfaces, users often interact with room-scale interfaces with their bodies (e.g., leaning against, sitting down, stepping on), thus the actuator must sustain such external force.
We considered inflatables actuators as it can withstand heavy objects, durable for unexpected interactions, and safe for users.

\subsubsection{Compact and Reconfigurable}
Room-scale shape-changing interfaces often require substantial space to install and deploy.
For temporary use or quick prototype, it is impractical to be permanently installed at one location.
Therefore we designed modular, compact, lightweight, and highly customizable units so that the user can quickly pick and place from storage to start prototyping.
% just like placing a tile on the floor.

\subsection{Mechanical Design}
\subsubsection{Plastic Tube and Constant Force Springs}
Here, we describe more detail about design and implementation.
The inflatable element is a 0.15 mm thickness vinyl tube that is 20 cm in diameter when inflated. We cut the tube to an appropriate length and heat-sealed both ends (and cut holes for valves later).  One end of the tube is rolled; the other is fixed to the base plate.
Each constant force spring (Dongguan Yongsheng Metal) has 1.5 m in length, 2 cm in width, and 0.8 kg force.
One end (flat side) of the spring is fixed to the base plate and the other end (rolled side) is fixed to the top end of the tube (Figure~\ref{fig:mechanical-design}).
Once fixed, we tape the spring to the side of the tube.

\begin{figure}[!h]
\centering
\includegraphics[width=3.4in]{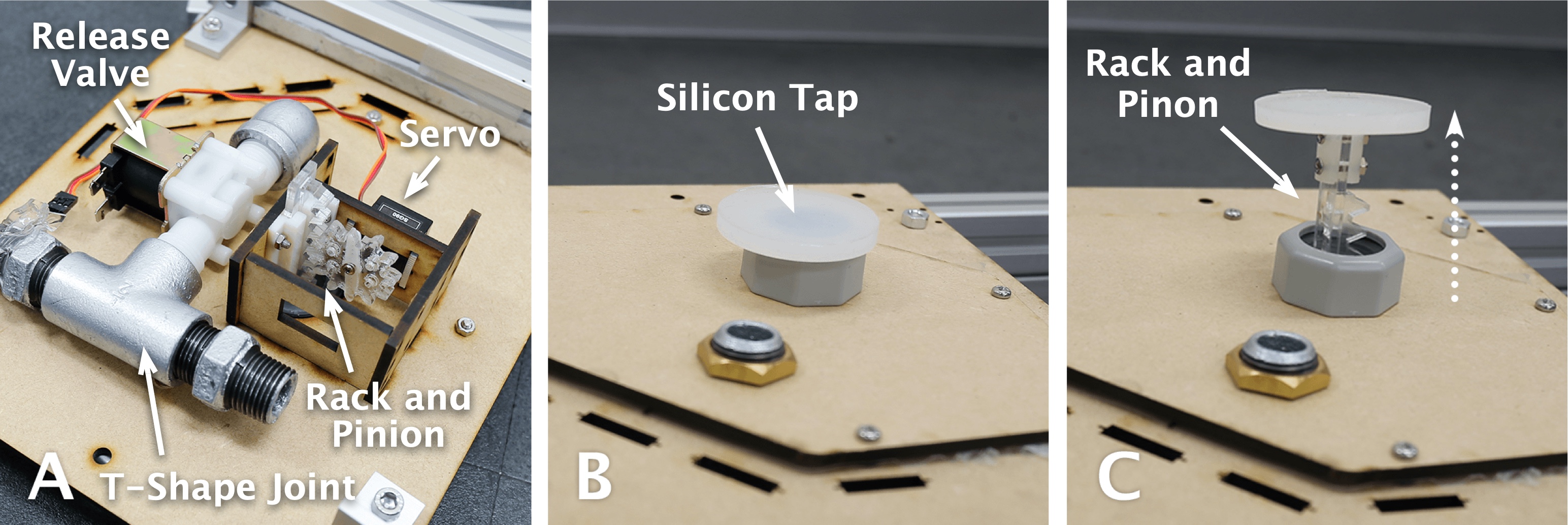}
\caption{The solenoid valve and the servo motor attached on the base plate (A). The servo motor, rack, and pinion mechanism actuate and move up the silicon tap to change from a closed state (B) to an open state (C) to release internal air.}
~\label{fig:system-valve}
\vspace{-0.6cm}
\end{figure}

\subsubsection{Base Plate and Valves}
The laser-cut base plate (plywood, 3 mm thickness, 22 cm x 22 cm size) has two holes for air supply and release respectively.
The supply hole is fitted with a 1.5 cm diameter solenoid valve (Ebowan Plastic Solenoid Valve DC 12V).
The release hole has a T-shaped silicon tap (Figure~\ref{fig:system-valve}B), which opens and closes (Figure~\ref{fig:system-valve}C) with a 3D printed rack and pinion gear mounted on a servo motor (TowerPro SG90).
This custom silicon tap has a large hole size (4 cm), enabling faster module retraction than common solenoid valves.
Assembly time for each module is approximately 15-30 minutes, and the cost of each actuator is less than 8 USD, including the solenoid valve (3 USD), contact force springs (0.5 USD), servo motor (2 USD), and other material costs.
In total, we fabricated 25 units for our prototype.

\subsubsection{Modular Design}
\begin{figure}[!h]
\centering
\includegraphics[width=3.4in]{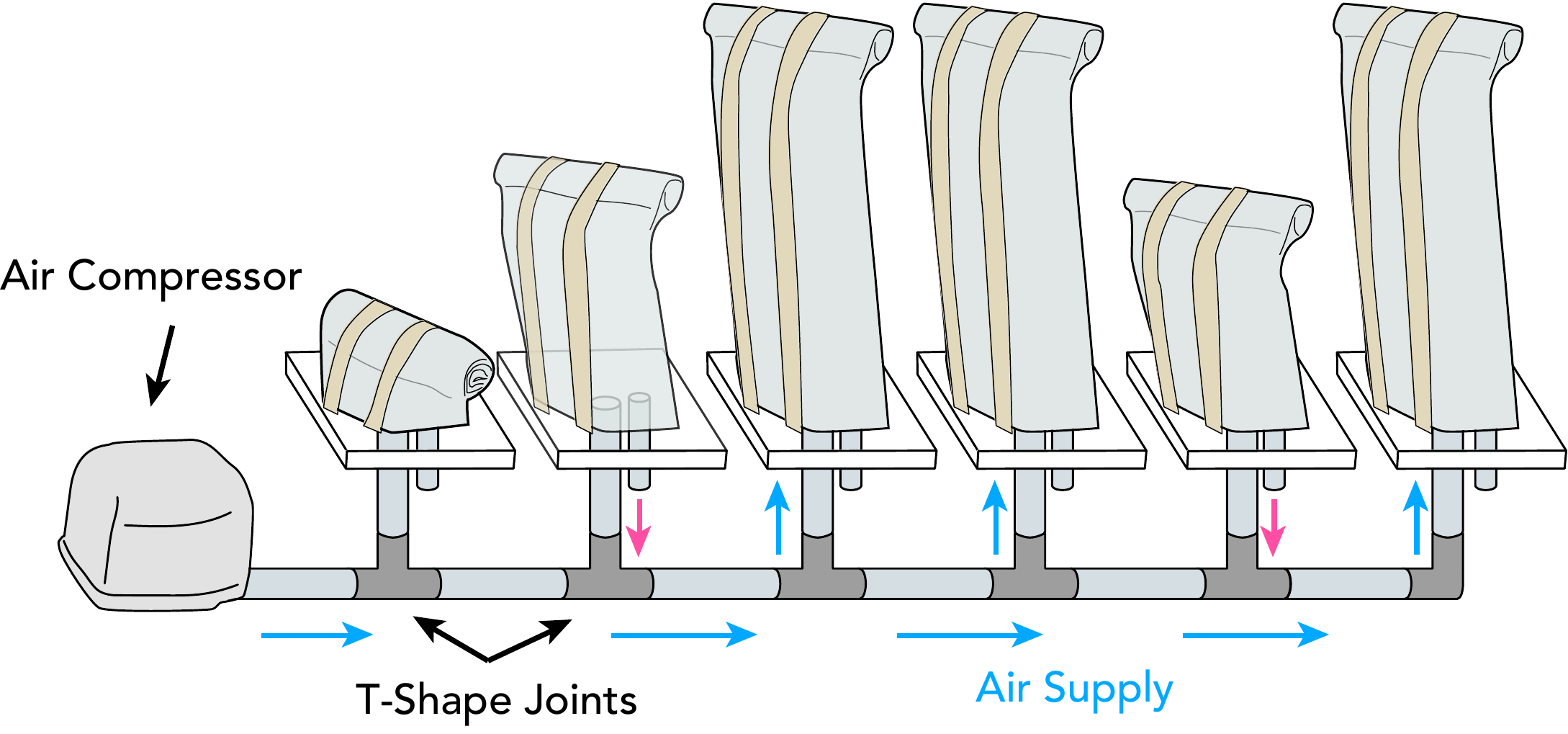}
\caption{Pneumatic control system for actuator arrays.}
~\label{fig:system-arrays}
\vspace{-0.6cm}
\end{figure}

Each actuator is modular and can connect with the air supply of neighboring actuators. 
Each solenoid air intake valve is connected to a T-fitting. 
Adjacent actuators are pneumatically connected with a silicon tube between the T-fittings (Figure~\ref{fig:system-arrays}). This way, an array of actuators is connected to a shared pressurized supply line.
Air compressors (Yasunaga Air Pump, AP-30P) pressurize the shared line. Each compressor provides up to 12.0kPa of pressure and supplies 30L of air volume per minute. 

\subsubsection{Control}
The height of each actuator is controlled by opening and closing a solenoid supply valve and a release tap actuated with a servo motor (Figure~\ref{fig:system-mechanism}).
To control each valve and servo motor, we used two 16 channel 12V relay modules (SainSmart) and 16 channel PWM servo motor drivers (PCA9685) respectively.
The servo motor driver and relay module are controlled from an Arduino Mega Micro-controller.
A Realsense SR300 depth camera mounted on the ceiling (or on a wall for horizontal extension) captures the current height of each actuator.
The software uses this sensor data to control the valves and to achieve the appropriate height.
Our client software written with OpenFrameworks displays the current height and control of each actuator and communicates with the master Arduino through a serial interface.

\begin{figure}[!h]
\centering
\includegraphics[width=3in]{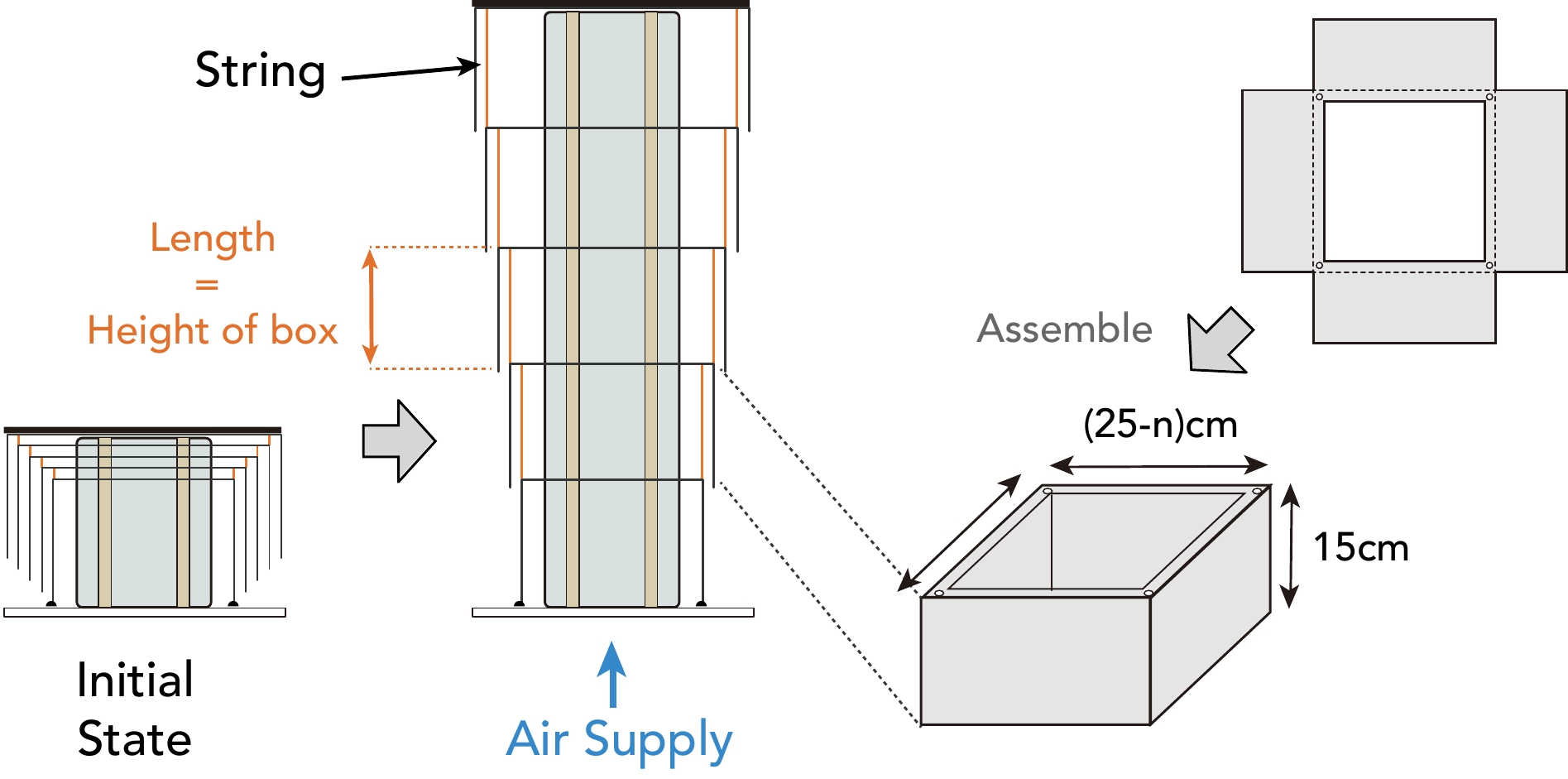}
\caption{Telescopic structure for a collapsible enclosure. Each square tube is connected with rubber bands (orange) to ensure they do not separate when extended.}
~\label{fig:system-telescopic}
% \vspace{-0.6cm}
\end{figure}

\subsubsection{Telescopic Enclosure}
When positioned vertically, the inflatable actuator has a rounded shape at the top, which is too unstable for stepping, standing or sitting.
Therefore, it is mounted inside a telescoped enclosure to enable the user to sit or step on it.
Each enclosure consists of ten concentric square tubes made of polycarbonate corrugated sheets (3mm thickness), one inside the other (Figure~\ref{fig:system-telescopic}).
We first laser cut the sheets (3mm thickness) and fold them into square tubes, then stack and connect them with Mobilon rubber bands (Nishinbo MB8064TA100G, 6mm thickness, 80mm folding diameter) to ensure they do not separate when extended.
Then, we mount and fix the telescoped enclosure to the base plate with cable ties. 
The top of the enclosure measures 30 cm x 30 cm and, when fully collapsed, 18 cm in height.
We measured the load-bearing capacity of the actuator by placing weights in 2kg increments on top of them. We found that the actuator and the telescopic enclosure can withstand at least 10kg.
With greater load, the telescoped enclosure starts to bend sideways, which made it difficult to accurately measure the maximum load. However, during use, we observed that multiple actuators stacked next to each other resisted bending and can withstand higher loads.

\subsection{Constructive Building Blocks}

\subsubsection{Vertical and Horizontal Orientations}
Due to the relatively compact size and small weight of the modules, various architectural surfaces can be considered for installation, such as placed on the floor or installed sideways from a wall.
% or horizontal surfaces like stairs hung from the ceiling 
% or free-standing pillar

\begin{figure}[!h]
\centering
\includegraphics[width=3.4in]{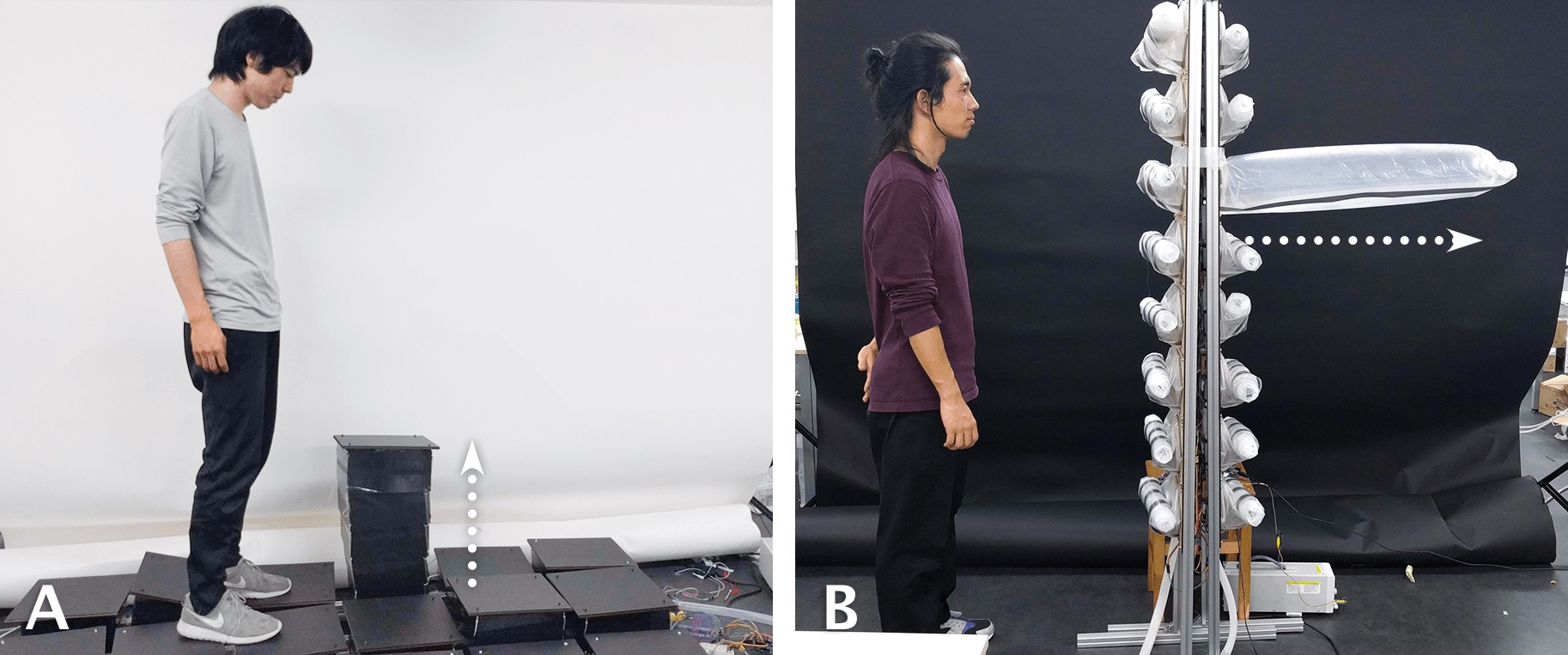}
\caption{Different orientation.}
~\label{fig:design-space-orientation}
\vspace{-0.6cm}
\end{figure}

\subsubsection{Different Sizes and Arrangements}
\begin{figure}[!h]
% \vspace{-0.2cm}
\centering
\includegraphics[width=3.4in]{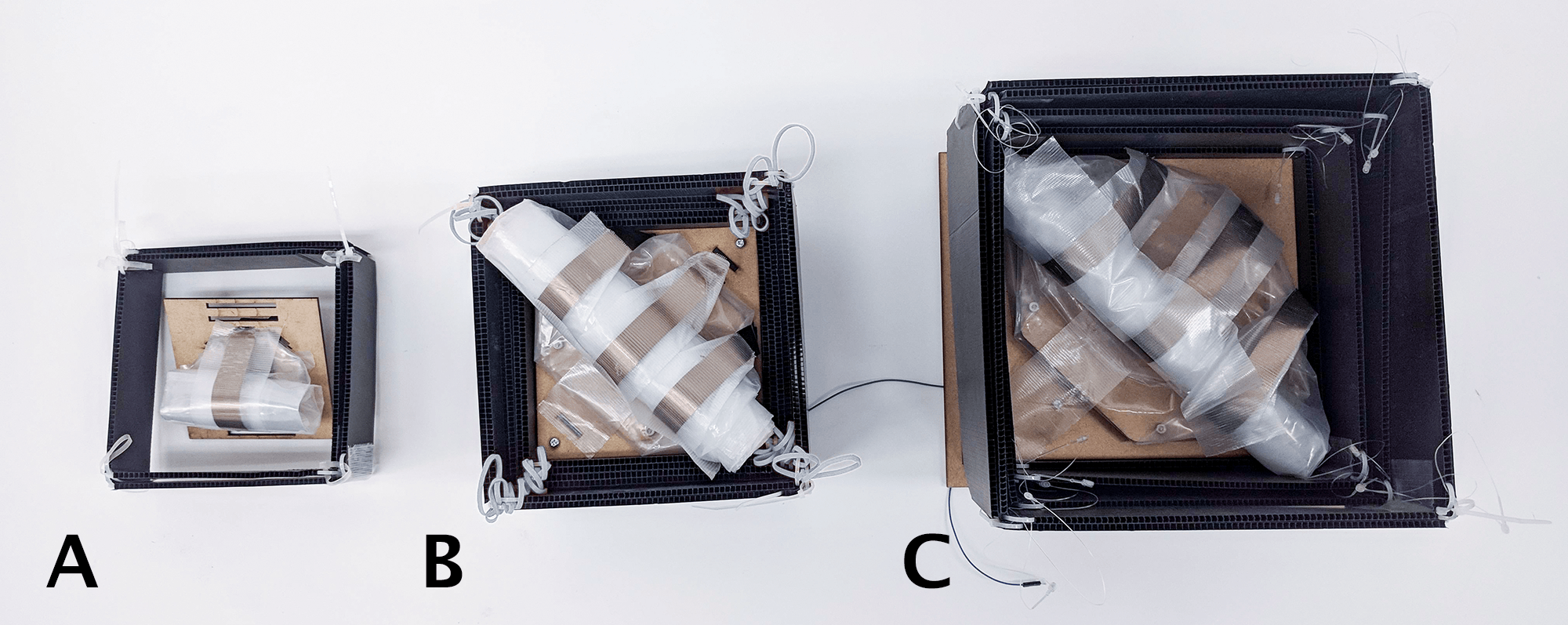}
\caption{Different actuator footprints: 10 cm, 20 cm, and 30 cm}
~\label{fig:design-space-size}
\vspace{-0.6cm}
\end{figure}

We also experimented with different sizes, trying footprints of 10 x 10 cm, 20 x 20 cm, and 30 x 30 cm, each of which uses a vinyl tube of 10 cm, 20 cm, and 27 cm in width respectively (Figure~\ref{fig:design-space-size}).
All are made with the same design, but the smaller designs require less force for the spring. We tested with 0.5kgf, 0.9kgf (0.3kgf x 3), 1.5kgf (0.5kgf x 3) for each size respectively.
We did not make the larger size than 30 x 30 cm as we were not able to find an off-the-shelf vinyl tube larger than 27 cm in width.

We observed that a smaller-sized actuator can respond faster, but was less stable under load.
Our final design uses the 30 cm x 30 cm size, as it has a higher load-bearing and a more stable structure, which is better suited for room-scale shape changes. 

% enables interesting applications such as reconfigurable furniture.
% large-scale interaction such as reconfigurable furniture
% a person to sit on a single actuator like a stool, a footprint of 30x30 cm was chosen. 

\begin{figure}[!h]
% \vspace{-0.2cm}
\centering
\includegraphics[width=3.4in]{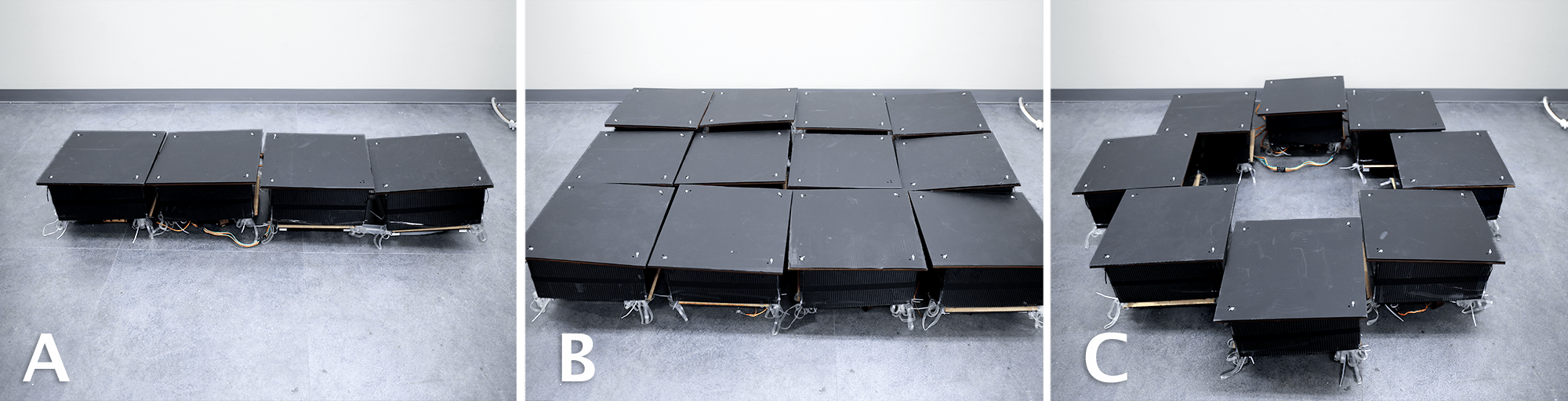}
\caption{Different actuator arrangements.}
~\label{fig:design-space-reconfigurability}
\vspace{-0.6cm}
\end{figure}

The actuators can be arranged in a grid, a line, or as individ- ual, loosely arranged units. Figure~\ref{fig:design-space-reconfigurability} depicts a few example configurations, which can also be reconfigured by the person installing them or the end user. An individual actuator can be picked up and moved, which may let the user handle it like a traditional piece of furniture, but could also open up new possibilities for interaction, such as actuators with a motorized base that reconfigures the arrangement according to the current use case.

\section{Prototyping Applications}
\begin{figure*}[!h]
\centering
\includegraphics[width=\textwidth]{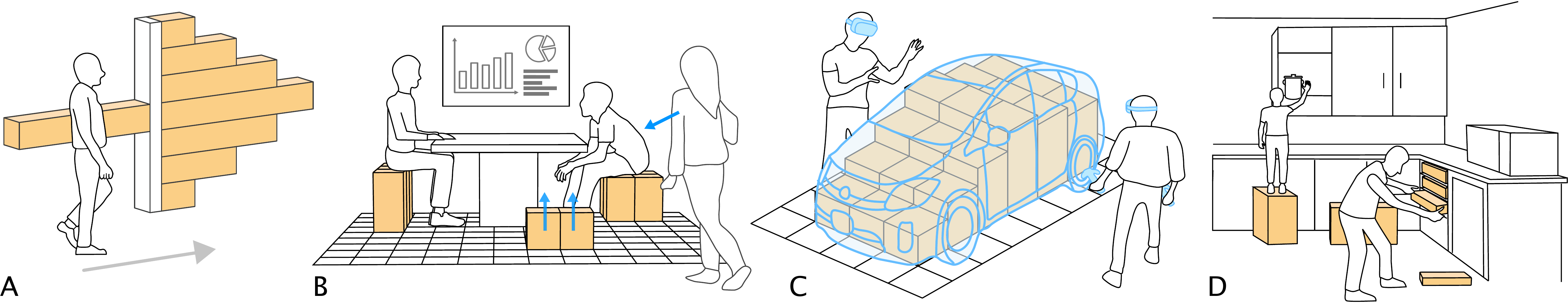}
\caption{Concept sketch of application scenarios, inspired by existing shape-changing interface research.
Orange part represents each LiftTile module.
(A) Wall-based display for wayfinding guidance.
(B) The room-scale shape display that becomes a chair, a table, or a bed.
(C) Large-scale haptics for virtual and augmented reality.
(D) Pick up and place for deployable pop-up structure.
}
~\label{fig:applications-sketch}
\vspace{-0.6cm}
\end{figure*}

To gauge how well \tool{} is suited to our goal of supporting rapid prototyping and user testing, we built several different example applications.
These were inspired by existing shape-changing interfaces, which we categorized into the following high-level application spaces: 1) {\it adaption:} adapting to users' needs and context~\cite{yao2013pneui, nakagaki2015lineform, roudaut2013morphees, gomes2013morephone, follmer2013inForm, gross2012architectural}, 2) {\it display:} communicating with the user by displaying information or data~\cite{follmer2013inForm, taher2015exploring, suzuki2018dynablock, goulthorpe2001aegis}, and 3) {\it haptics}: providing haptic feedback synchronized with visual information~\cite{siu2018shapeShift, teng2018pupop, delazio2018force, iwata2001project}.
Figure~\ref{fig:applications-sketch} depicts a concept sketch of applications resulting from an initial brainstorming. The following example applications were all chosen to leverage large-scale shape change and built in less than a day.

\subsubsection*{Adaptive Floor Furniture}
The first application is a reconfigurable floor that creates adaptive furniture, inspired by an adaptive dynamic table (e.g., TRANSFORM~\cite{vink2015transform}) and dynamic physical affordances (e.g., inFORM~\cite{follmer2013inForm}).
We created a shape-changing floor by arranging 25 modules in a 5 x 5 grid. Each unit is individually actuated to provide a chair, a table, and a bed (Figure~\ref{fig:cover}).

\subsubsection*{Shape-changing Wall}
By leveraging the block's compact form factor, we also prototyped a horizontal shape-changing wall to adapt to user needs and display information.
The wall becomes a space separator to make a temporary meeting space, while it can also act as public signage by displaying an arrow shape to guide a passer-by.
% By stacking multiple horizontal displays along a wall, they can render more expressive information, such as forming a pictograph through negative space.

\begin{figure}[!h]
\centering
\includegraphics[width=3.4in]{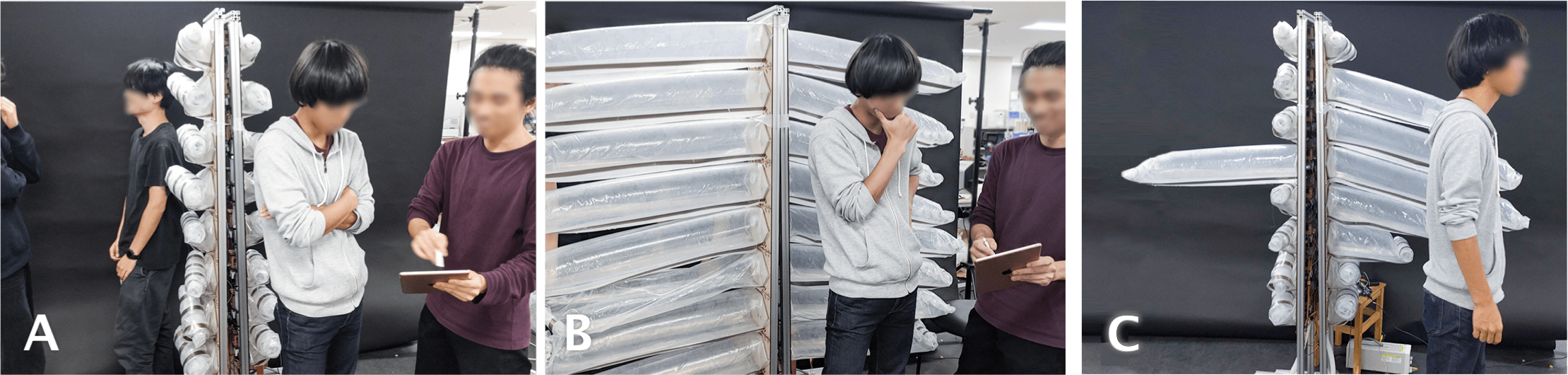}
\caption{Adaptive space separation. The wall creates a temporary meeting space by separating the space, and changes shape to display an arrow.}
~\label{fig:applications-separation}
\vspace{-0.6cm}
\end{figure}

\subsubsection*{Deployable Pop-Up Structure}
As the actuators are compact compared to traditional furniture and walls, they can be deployed in temporary use cases such as festivals, trade shows, concerts, and disaster relief.
In these applications, the actuators are shipped in a compact box and laid out in an empty room, similar to tiles.
After connecting them to a compressor and control computer, the operator selects a use case and the blocks render a layout of chairs, beds, tables, and partitions.
% This arrangement could be pre-programmed or it could be defined by the operator on-site by specifying what shapes to extrude. After the event, the actuators contract into compact tiles and can be stored for future use.

\begin{figure}[!h]
\centering
\includegraphics[width=3.4in]{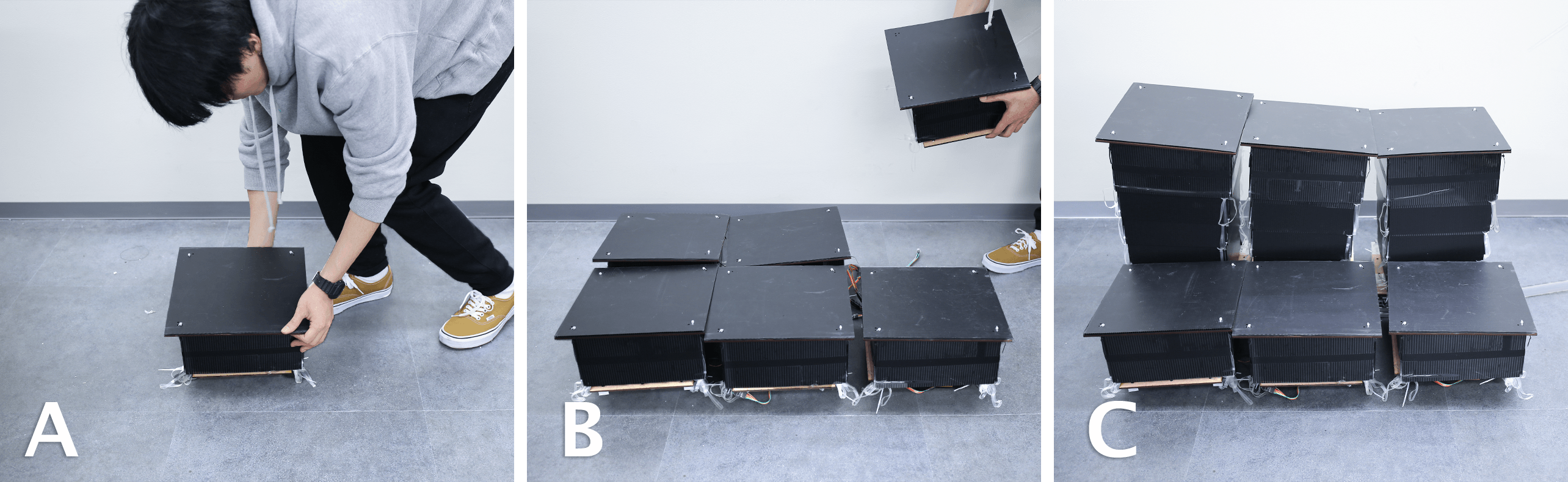}
\caption{Modular design to construct different geometries.}
~\label{fig:system-modular}
\vspace{-0.6cm}
\end{figure}

\subsubsection*{Landscape Haptics for VR}
While hand-sized shape displays have rendered haptic sensations for VR~\cite{siu2018shapeShift}), we propose to render furniture-size props, walls, and game elements that users can feel through walking and touching while experiencing immersive VR scenes.
\changes{Similar to the applications proposed by HapticTurk~\cite{cheng2014haptic}, TurkDeck~\cite{cheng2015turkdeck}, and TilePop~\cite{teng2019tilepop}, In this application, the haptics increases the feeling of presence in games by providing room-scale dynamic haptic feedback.}
Also, this can aid the design process by rendering full-scale object mock-ups of large objects like a car.

\begin{figure}[!h]
\centering
\includegraphics[width=3.4in]{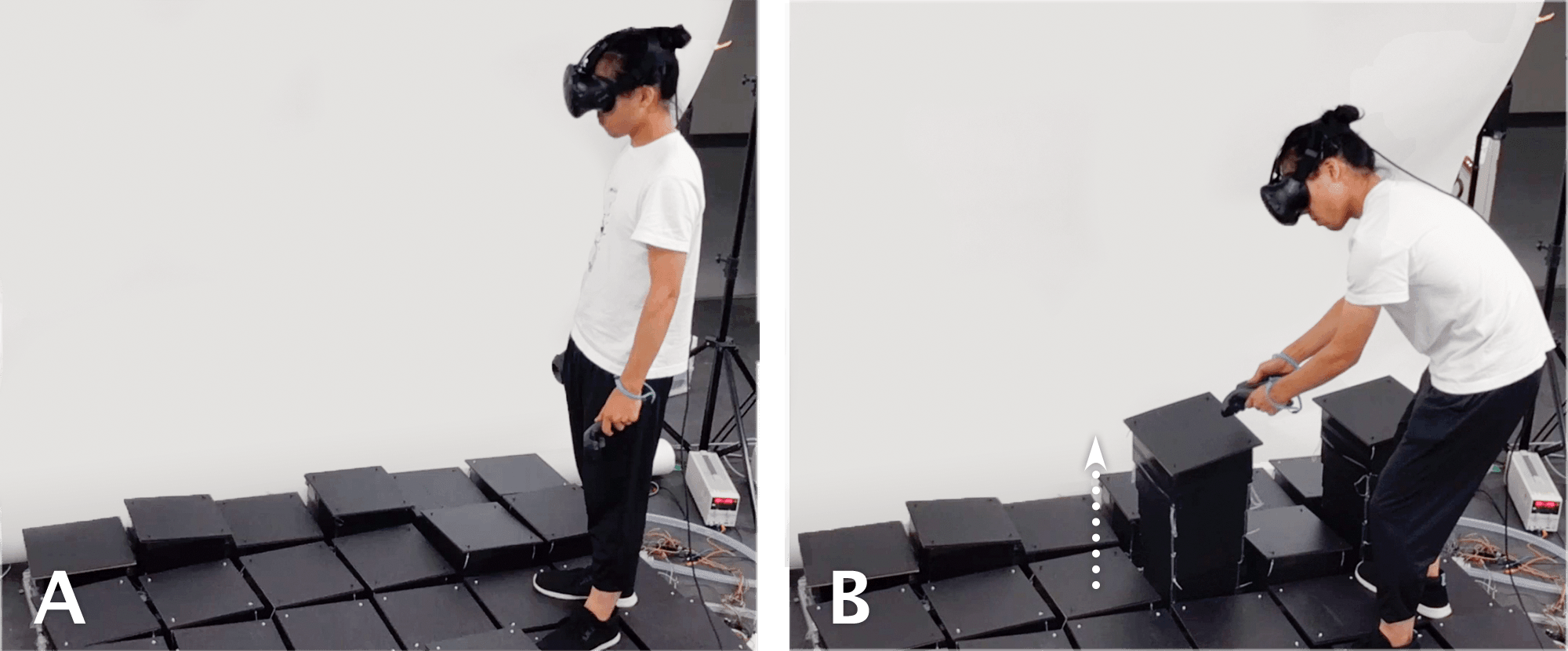}
\caption{Room-scale dynamic haptics for VR.}
~\label{fig:applications-haptics}
% \vspace{-0.6cm}
\end{figure}

\subsubsection*{Dynamic Data Physicalization}
Inspired by related work in dynamic data physicalization~\cite{Jansen:2015:OCD:2702123.2702180, taher2015exploring}, this exhibit enables for a tangible exploration of data.
On the floor, a timeline of the last 2,000 years is shown, and visitors move an actuator module along the timeline. The actuator dynamically extends or retracts to represent the average body height of a person of the same age at the given period. For example, a 10 years old girl can see and feel how children 100 years ago were of a different average height.
% We made this for a data discovery room in a science museum.

\section{Limitations and Future Work}
During the process of prototyping example applications, we confirmed that the modularity and simplicity of \tool{} are well-suited for rapid prototyping, but we also discovered a number of limitations that we plan to address in future work. 

\subsubsection*{Load Tolerance}
Although the pneumatic actuator can withstand heavier load than 10 kg, our outer enclosure is insufficiently stable and robust, as discussed in the system section (see Telescopic Enclosure).
We added the fabricated enclosure to stabilize the top, but as it is manually fabricated out of plastic cardboard, the structure becomes unstable when loaded with heavy objects.
\changes{A user can sit on a single actuator, but has to actively maintain balance while sitting, which is not ideal even for prototyping scenarios. }
A more precisely manufactured telescoping structure would greatly increase the load tolerance.
\changes{An alternative possible solution would be to use a bi-stable or mono-stable extendable structure such as tuneable mechanical truss~\cite{yasuda2017origami}, origami-based enclosure~\cite{zhai2018origami}, and {\it coilable} masts~\cite{joosten2007preliminary}, inspired by deployable structures explored for space applications~\cite{jensen2001arm} (e.g., deployable antennas and masts.)
Finally, combining multiple reel-based tubes inside each block can make them more stable. This design also allows the larger blocks than 30 cm. 
One potential challenge is that each tube may inflate at a different speed even if air comes from the same supply hole, but this can be addressed by using pressure sensor or height sensor to control the amount of air to be supplied.
We will also investigate this as future work}

\subsubsection*{Speed of Transformation}
One current limitation is the slow speed of transformation. In our current setting, \changes{it takes 16 seconds to fully extend a module and 4 seconds to fully retract it}.
There are some potential solutions, such as using a stronger air compressor and larger valves, but the speed of transformation of large inflatables is generally slower than electromechanical actuation. 
It is therefore difficult to prototype applications that require fast shape transformation, such as VR games. Interactions similar to the ones proposed in related work like TurkDeck~\cite{cheng2015turkdeck} and TilePop~\cite{teng2019tilepop} can be utilized to maintain a sense of presence for the user even with slow transformations.

\subsubsection*{Lack of Interactivity}
This paper we focused primarily on the {\it output} of shape-changing interfaces and less on sensing user input.
Adding input sensing modalities such as embedded touch sensing would greatly increase the interactive capabilities.
\changes{For future work, incorporating the sensing methods proposed by~\cite{swaminathan2019input} (e.g., acoustic, capacitive, and pressure sensing) should be explored to improve the interactivity of the current prototype.}
We are also interested in using a dynamic appearance (e.g., color changing with LEDs) to provide different affordances.

\subsubsection*{Different Height Tracking}
IR photo reflectors or pressure sensors could replace the external camera for a more robust internal height sensing.
We observe that sometimes two similar tubes extend at a different rate, although inflating from the same pump.
More precise and embedded sensing can achieve more consistent speed and control.

\subsubsection*{Distributed Control}
Currently, we are controlling each module through a central Arduino board, which requires a cumbersome electric connection with wires. We are interested in improving the modularity of each unit by using a Wifi chip module (e.g., ESP8266) to control wireless communication. This would also allow easy mapping between modules and the software interface (currently, we need to manually map at the beginning.) 
We are also interested in adding mobility in each unit to allow for autonomous and distributed room-scale shape-changing interfaces. 

\subsubsection*{User Evaluation}
All our prototypes are created by us, and we have not formally evaluated the system with a user study. As a next step, we plan a workshop with researchers and designers (e.g., interior designers and architects) to explore and prototype, similar to ~\cite{hardy2015shapeclip}.
We expect a formal evaluation to provide insights into how this tool supports design ideation.

\section{Conclusion}
This paper introduces constructive building blocks for prototyping room-scale shape-changing interfaces.
To this end, we designed a modular inflatable actuator that is highly extendable (from 15cm to 150cm), low-cost (8 USD), lightweight (10 kg), compact (15 cm), and strong (e.g., can support 10 kg weight) making it well-suited for prototyping room-scale shape transformations.
Moreover, our modular and reconfigurable design allows researchers and designers to quickly construct different geometries and to investigate various applications.
In this paper, we contributed to the design and implementation of highly extendable inflatable actuators, and demonstrates a range of scenarios that can leverage this modular building block.
We plan to evaluate the potential of our system by inviting designers to use our blocks in their own designs and see whether and/or how our building blocks enhances their ideation process.

% design space for room-scale shape displays, implement a prototype system and ideate applications with three domain experts. Based on their feedback, we propose a number of example applications and envision scenarios for reconfigurable architectural spaces, temporary deployable pop-up structures, and haptic feedback for VR. 

% \tool{}, a modular and reconfigurable room-scale shape display. To achieve actuation of furniture scale objects, we design and implement a retractable inflatable actuator, which enables a compact, lightweight, modular structure. We propose a design space for room-scale shape displays, implement a prototype system and ideate applications with three domain experts. Based on their feedback, we propose a number of example applications and envision scenarios for reconfigurable architectural spaces, temporary deployable pop-up structures, and haptic feedback for VR. 

% In this paper, we introduced Dynaroom, a system to explore room-scale Human-Architecture Interaction. Based on an array of low-cost, light-weight, and high load-bearing pneumatic linear actuators, we demonstrate how floors and walls can turn into dynamic interactive surfaces. These room-scale shape displays render partitions and furniture on demand, display information physically and provide dynamic haptic feedback. We envision applications scenarios for reconfigurable architectural spaces, temporary deployable pop-up structures, and Virtual Reality. 
\section{Acknowledgements}
We thank Shohei Takei for helpful suggestions.
This research was supported by the JST ERATO Grant Number JPMJER1501 and the Nakajima Foundation.

\balance
\bibliographystyle{ACM-Reference-Format}
\bibliography{references}

\end{document}